# A Fiber-Optic Platform for Cryogenic Laser-Pump Neutron-Probe Magneto-Optical Measurements

Lazar L. Kish[1], Andrei T. Savici[2], Tyler K. White[2], Vlad N. Sedov[2], Andrey A. Podlesnyak[2], Matthias D. Frontzek[2], Barry L. Winn[2], and Igor A. Zaliznyak[1]

1. Condensed Matter Physics and Materials Science Division, Brookhaven National Laboratory, Upton, NY, USA

2. Neutron Scattering Division, Oak Ridge National Laboratory, Oak Ridge, TN, USA

*zaliznyak@bnl.gov*

***Abstract*— We report a fiber-optic laser-pump neutron-probe method for momentum- and frequency-resolved magneto-optical measurements. The technique uses inelastic neutron scattering to measure time-dependent two-point magnetization correlations, S(Q,$\omega$), after resonant photoexcitation at cryogenic temperatures. We demonstrate it by measuring nonequilibrium thermalization dynamics in Yb-based quantum magnets.**

## I. Introduction

In magneto-optical materials, incident photons couple to electronic states that are linked to atomic magnetic moments. A major goal in the field is to develop optically interfaced solid-state spin platforms for quantum information storage, processing, and networking. Some particularly attractive examples are optical insulators containing rare-earth ions because their partially filled $4f$ shells are shielded by outer electronic orbitals, producing narrow optical transitions, long-lived spin states, and weak coupling to many environmental perturbations. In rare-earth Kramers ions, the crystal-field ground state often forms a well-isolated magnetic doublet that maps onto an effective quantum spin-1/2, or equivalently, a qubit. These properties have motivated rare-earth materials for solid-state quantum memories and long-range entanglement distribution [1].

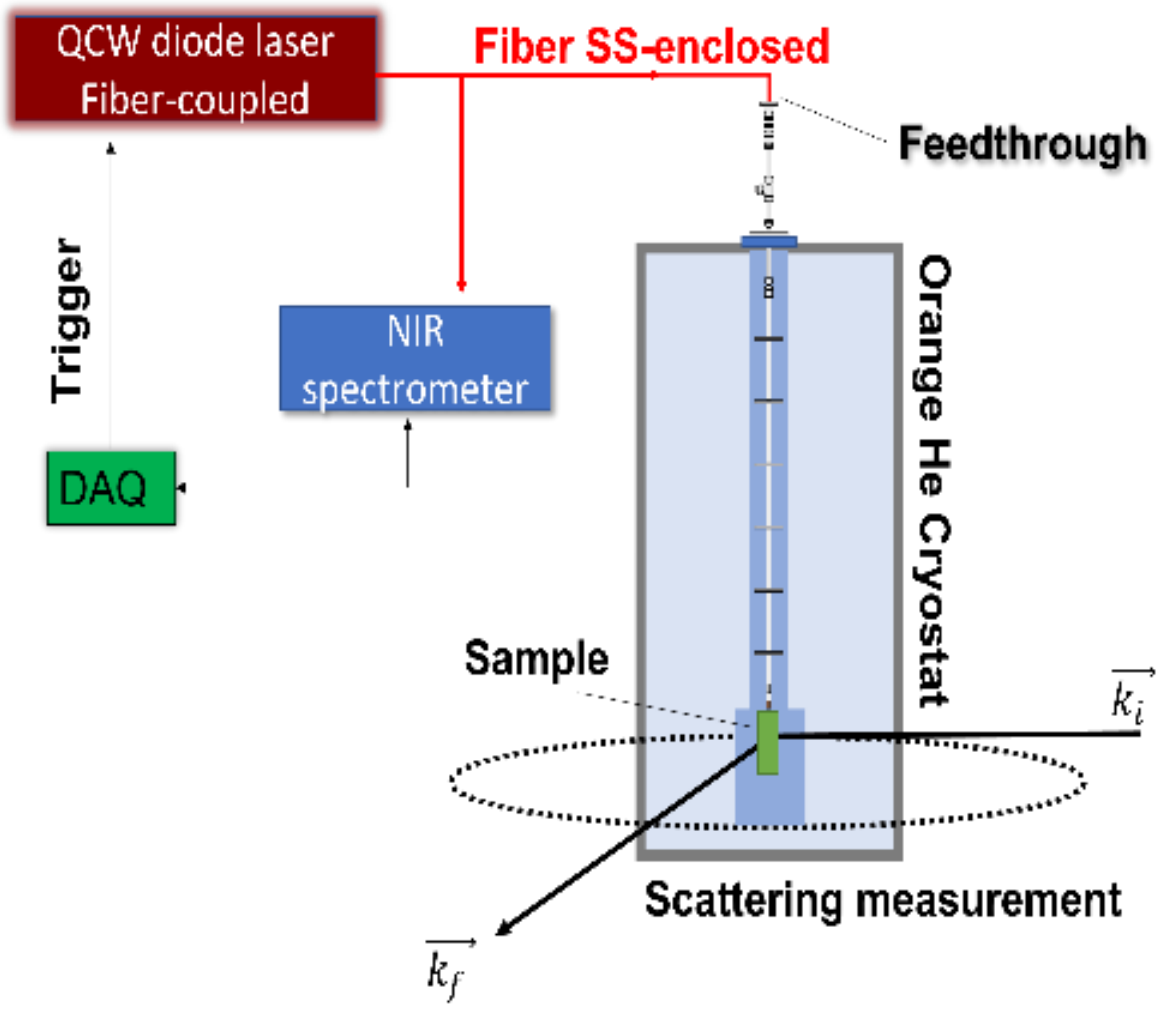

*Fig. 1. Laser-pumped neutron-probe platform schematic*

Much of the focus in this regard has been on isolated-ion systems, where optical transitions and coherence times can be characterized by conventional optical spectroscopy. In concentrated magnets however, the same optically active ions also carry interacting magnetic moments, where resonant excitation can perturb entangled many-body ground-states and collective quantum excitations of many coupled ions. The interaction of light with these collective magnetic states is poorly understood, in part because few probes can directly resolve the magnetic response after optical excitation in both momentum and frequency.

Our collaboration between Brookhaven National Laboratory and Oak Ridge National Laboratory has developed a new fiberoptic platform [2] for nonequilibrium measurements of the momentum and energy-resolved magnetic correlation functions using time-resolved neutron scattering to capture nonequilibrium dynamics under laser photoexcitation at cryogenic temperatures. Here we report the development of this setup and describe its initial deployment at the Spallation Neutron Source and High Flux Isotope Reactor. The accompanying presentation will demonstrate the first measurements employing this platform.

## II. Experimental Platform

### A. *Cryogenic sample environment with in-situ photoexcitation carried by optical fiber*

Fig. 1. Shows an abstracted schematic of the experimental platform with its various components. The sample to be measured is mounted in an “Orange” He-4 cryostat standard in neutron scattering facilities, which can reach a base temperature of ~1.6K without laser excitation. The insert is shown mounted in the cryostat in Figure 2A. These temperatures are often necessary in order to depopulate thermal excitations above the energy scale of quantum dynamics, depending on the materials system being measured. A custom sample-stick carries seven multimode optical fibers (1+6 bundle) into the cryostat which allow us to propagate laser light to the sample through the central fiber, with the rest used to measure backscattered light for complementary optical measurements. The cable enters the cryostat through a KF-40 flange at the top of the stick and is coupled on the ambient side to a fiber-coupled NIR diode lasers. Near the sample position, the laser light exits the fiber, passing through a fixed-focus collimation package which expands the beam to a waist-size of 2.5 mm (see inset of Fig. 2A). For the purposes of our initial measurements, we used high-power laser diodes which output light at 940 nm (3W) or 975 nm (1 W), wavelengths resonant with the crystal-field-split F7/2→ F5/2 transitions of $Yb^{3+}$ in perovskite and garnet versions of the Yb-Al-O system, our initial test materials. However, the fiber is versatile for use with many fiber-coupled light sources in the IR-NIR wavelength range. The diodes are temperature-controlled and driven by a current-source and can be modulated with a 100 kHz bandwidth. For our initial application, the modulation is a series of square pulses triggered by a TTL timing voltage signal synchronized with the neutron pulses of the Spallation Neutron Source and the time structure of the data acquisition system of the instrument. The 6 pickup optical fibers carry light backscattered from the sample to measurement equipment outside the cryostat, which can be an NIR spectrometer or high-bandwidth photodiode for complementary optical measurements or diagnostics, depending on the needs of the experiment. The time dependent optical signal can be logged with experiment metadata for comparison with neutron intensities during data analysis.

### B. *Time-resolved neutron scattering under photoexcitation*

Neutron scattering beamlines which can employ this platform have been equipped with event-mode data collection of neutrons, tagging each measured neutron with an individual time stamp which can be referenced to a time delay away from a laser pulse [5]. At the Spallation Neutron Source, neutrons come with a frequency of 60 Hz. Time-resolved measurements can then either be achieved stroboscopically, by altering the time-delay between the laser and neutron pulses with a time-resolution similar to the neutron pulse-width (20~50 microseconds), or built up on a slower timescale by pulsing the laser once every few neutron pulses ('frames') and measuring subsequent neutron frames to build a time-dependence. In contrast, at the High-Flux Isotope Reactor, a constant, high-intensity beam of neutrons irradiates the sample for continuous data collection, and time-resolved data can be arbitrarily resampled in analysis.

Instruments such as the Hybrid Spectrometer (HYSPEC)[3] which are designed for inelastic neutron spectroscopy provide an energy-resolved, momentum-resolved neutron spectrum which can be time-sliced to produce spectral maps at various time delays from laser excitation. Meanwhile, instruments designed for neutron diffraction measurements produce a momentum-resolved map of neutron intensity which can be filtered by time-delay away from laser excitation to produce a time-dependence of the magnetic or crystal structure.

### C. *First experiments*

This setup has been successfully deployed in experiments at both the Spallation Neutron Source and the High-Flux Isotope reactor across three different beamlines, measuring neutron scattering from Yb-based quantum materials [2] under time-dependent laser excitation. Fig. 2A shows the setup mounted on the $WAND^2$ diffractometer at the High-Flux Isotope Reactor. Figs. 2 B, C show measurements on the HYSPEC instrument at the Spallation Neutron Source of quantum spinon excitations in the spin-chain material $YbFeO_3$, which undergoes a magnetic phase transition near 8K. Figs. 2 D-G demonstrate the time evolution of such spectra under periodic laser pumping every 8 neutron pulses (spaced 16 ms apart), integrated along the wave-vector direction and taken as a difference over the last neutron pulse before the next firing of the laser. Counterintuitively, the nature of the spectral shift indicates a cooling of the Yb spins by each laser pulse. This could be a sign of spinon-assisted laser cooling processes or Yb-Fe entropy transfer during subsystem thermalization. More details about our initial results from these measurements will be discussed in the accompanying presentation.

## III. References

[1] [1] D. D. Awschalom, R. Hanson, J. Wrachtrup, B. B. Zhou, *Quantum technologies with optically interfaced solid-state spins*, Nature Photonics **12**, 516 (2018).

[2] L. L. Kish, A. Weichselbaum, D. M. Pajerowski, A. T. Savici, A. Podlesnyak, L. Vasylechko, A. Tsvelik, R. Konik, I. A. Zaliznyak. "High-temperature quantum coherence of spinons in a rare-earth spin-chain." Nature Communications. (2025) 16:6594

[3] I. A. Zaliznyak, A. T. Savici, V. Garlea, *et al.* 2017, "Polarized neutron scattering on HYSPEC: the HYbrid SPECtrometer at SNS." J. Phys.: Conf. Ser. 862 012030

[4] C,.Hua, B. L. Winn, C. Sarkis, *et al.* Nat Commun **17**, 3535 (2026)

[5] Peter F. Peterson, et al., "Event-based processing of neutron scattering data, Nuclear Instruments and Methods in Physics Research Section A: Accelerators, Spectrometers", Detectors and Associated Equipment, Volume 803, 2015, Pages 24-28, ISSN 0168-9002,

[6] L. L. Kish, T. K. White, V. N. Sedov, M. D. Frontzek, A. Podlesnyak, B. L. Winn, I. A. Zaliznyak. "A fiberoptic platform for laser-pumped neutron-probed magnetooptic measurements at cryogenic temperatures." Unpublished.

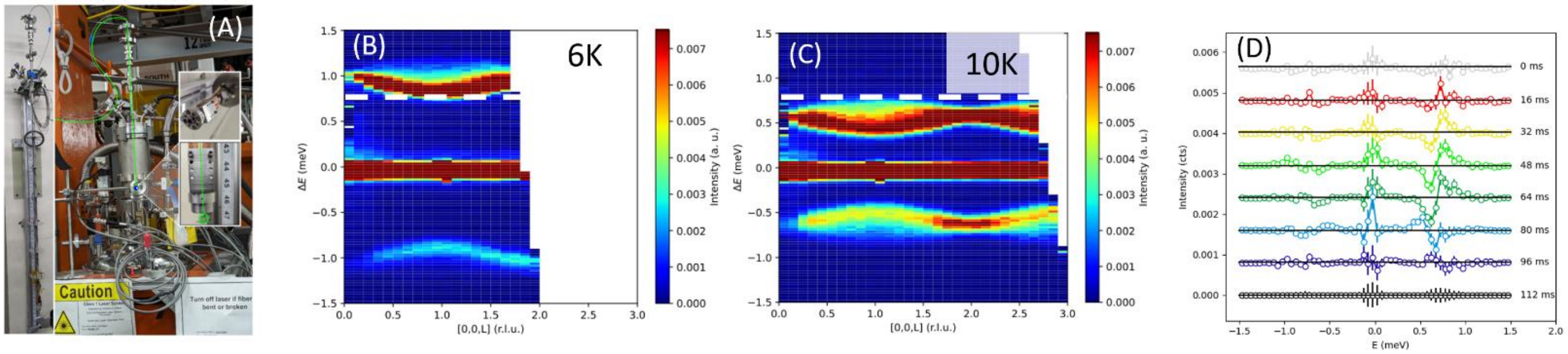

*Fig. 2. Laser-pumped neutron-probed measurements on $YbFeO_3$. (A) Picture of fiberoptic platform mounted on cryostat (B, C) Magnetic spectrum at $YbFeO_3$ spin-reorientation transition (D) Time evolution of magnetic spectrum integrated along wave-vector between pulses*

Work at Brookhaven was supported by Office of Basic Energy Sciences (BES), Division of Materials Sciences and Engineering, U. S. Department of Energy (DOE), under contract DE-SC0012704.